\documentclass[prb,preprint,showpacs,superscriptaddress,
preprintnumbers,amsmath,amssymb,nofootinbb]{revtex4-1}
\usepackage{graphicx}
\usepackage{dcolumn}
\usepackage{multirow}

\usepackage{amsmath}
\usepackage{subfigure}
\usepackage{braket}
\usepackage{color}
\usepackage{epstopdf}
\usepackage[normalem]{ulem}

\newcommand{\MC}[1]{\textcolor{blue}{{#1}}}

\begin{document}
\title{First-principles molecular dynamics study of deuterium diffusion in liquid tin}

  \author{Xiaohui Liu}
\affiliation{CAS Key Laboratory of Quantum Information, University of Science and
  Technology of China, Hefei, Anhui, 230026, People's Republic of China}
\affiliation{Synergetic Innovation Center of Quantum Information and Quantum
  Physics, University of Science and Technology of China, Hefei, 230026, China}
  \author{Daye Zheng\footnote{Xiaohui Liu and Daye Zheng contributed equally to this work.}}
\affiliation{CAS Key Laboratory of Quantum Information, University of Science and
  Technology of China, Hefei, Anhui, 230026, People's Republic of China}
\affiliation{Synergetic Innovation Center of Quantum Information and Quantum
  Physics, University of Science and Technology of China, Hefei, 230026, China}

\author{Xinguo Ren}
\affiliation{CAS Key Laboratory of Quantum Information, University of Science and
  Technology of China, Hefei, Anhui, 230026,  People's Republic of China}
\affiliation{Synergetic Innovation Center of Quantum Information and Quantum
  Physics, University of Science and Technology of China, Hefei, 230026,
  China}
\author{Lixin He}
\email{helx@ustc.edu.cn}
\affiliation{CAS Key Laboratory of Quantum Information, University of Science and
  Technology of China, Hefei, Anhui, 230026,  People's Republic of China}
\affiliation{Synergetic Innovation Center of Quantum Information and Quantum
  Physics, University of Science and Technology of China, Hefei, 230026, China}
  \author{Mohan Chen}
\email{mohan.chen@temple.edu}
\affiliation{Department of Physics, Temple University, Philadelphia, Pennsylvania 19122, USA}
\date{\today }

\begin{abstract}
Understanding the retention of hydrogen isotopes in liquid metals, such as lithium and tin, is
of great importance in designing a liquid plasma-facing component in fusion reactors.
However, experimental diffusivity data of hydrogen isotopes in liquid metals are still limited or controversial.
We employ first-principles molecular dynamics simulations to predict
diffusion coefficients of deuterium in liquid tin at temperatures ranging from 573 to 1673 K.
Our simulations indicate faster diffusion of deuterium in liquid tin than the self-diffusivity of tin.
In addition, we find that the structural and dynamic properties of tin are insensitive to the inserted deuterium
at temperatures and concentrations considered.
We also observe that tin and deuterium do not form stable solid compounds.
These predicted results from simulations enable us to have a better understanding of the retention of hydrogen isotopes in liquid tin.

\end{abstract}

\maketitle

\section{INTRODUCTION}
Designing reactors that can generate fusion energy as
a viable energy source has been a great challenge for decades.
One of the challenging issues is to build plasma-facing components
that survive intense particle bombardments present in the
harsh environment of a fusion reactor.
In this regard, solid plasma-facing materials unavoidably
suffer from erosion when they are exposed to high fluxes of particles,\cite{lipschultz2012divertor,pitts2013full,van2014effect}
which may also lead to performance degradation of the plasma-facing components.
Recently, several experiments were conducted to study liquid metals,
which own a series of attractive properties, as alternative plasma-facing materials.
\cite{coenen14,kaita07,mazzitelli2010review,majeski13,fiflis2014wetting,tritz14,schmitt15,abrams2015modeling,16NF-Abrams,morgan15,16L-Eden}
In particular, lithium (Li) is a low-Z metal that can be tolerated in plasma and has been widely used in these recent experiments.
For instance, the performance of plasma in fusion reactors has been improved by using liquid lithium.
\cite{kaita07,mazzitelli2010review,majeski13,fiflis2014wetting,tritz14,schmitt15,abrams2015modeling}
Moreover, recent studies found liquid lithium with inserted deuterium atoms
can transform to solid lithium deuteride at high temperatures, resulting in largely suppressed sputtering yields\cite{16NF-Abrams,16NF-Chen}.
These studies also suggested that the operating temperature of lithium as plasma-facing material can be higher than previously thought.
However, a significant change of thermal properties during the liquid-to-solid phase transition is expected,
which is in general difficult to predict.

In order to be easily melted and then operated as plasma-facing materials,
liquid metals should have a low melting point relative to the solid counterparts.
Tin has a slightly higher melting point of 505 K\cite{12JAP-Weir}
than lithium (453 K),\cite{83B-Boehler} and is emerging as an alternative
liquid metal of plasma-facing material.
For example, a recent experiment by van Eden et al.\cite{16L-Eden}
showed that liquid tin possesses vapor-shielding effect under ion bombardments
at very high temperatures,
resulting in a reduced heat flux on the liquid tin surface.
Liquid tin owns some substantial advantages over liquid lithium.
For instance, liquid tin can sustain higher surface temperatures because of
its lower vapor pressures and material losses compared to liquid lithium.\cite{02FES-Brooks,03-Coventry,04-Coventry}
Moreover, the retention of hydrogen in liquid tin was found to be much smaller than that of liquid lithium.\cite{loureiro2016deuterium}
Tin also does not combust in the presence of sufficient water vapor, which is a significant advantage in terms of safety.

Hydrogen isotopes are fuels and their retention and recycling issues
in an operated fusion reactor need to be thoroughly understood.\cite{causey2002hydrogen}
One of the unsolved issues is that the diffusion coefficients of hydrogen isotopes
in liquid metals have not yet been well documented. Particularly, the diffusivities of hydrogen isotopes in
liquid lithium are still controversial.\cite{moriyama1992transport,fukada2005hydrogen}
In addition, to the best of our knowledge, the diffusivity of hydrogen isotopes in liquid tin still lacks.
It is difficult to obtain accurate diffusivity data in experiments, in part due to the existence of
impurities in liquid samples.
For instance, experiments found oxygen atoms form oxides with
both lithium\cite{fiflis2014wetting} and tin\cite{03-Coventry} and
suggested the impacts of these impurities on the properties of liquid metals cannot be ignored.
Notably, the diffusivities of hydrogen isotopes in liquid metals are able to be predicted from computational simulations,
and the structural and dynamic properties of liquid metals can also be obtained.

With the fast development of computational methods and resources in the last few decades,
simulations have become an indispensable tool in materials science.
Computational methods are well suitable to study properties of systems with or without impurities at
different external conditions and provide further insights or even predict new properties in support of experiments.
Recent computational works of liquid lithium\cite{chen13,vella14,Chen15,16NF-Chen,chen16,17B-Vella}
have been proven to be reliable in explaining experiments and predicting structural and dynamic properties of liquid metals.
For example, a first-principles molecular dynamics (FPMD) study \cite{16NF-Chen}
confirmed the suspected formation of lithium deuteride in experiments.\cite{abrams2015modeling,16NF-Abrams}
By using atomistic simulation methods based on force fields,
several thousands of particles were simulated up to a few nanoseconds
in order to obtain temperature-dependent vapor-liquid surface tensions and viscosities of liquid lithium\cite{vella14,Chen15}
and liquid tin.\cite{17B-Vella}

However, force fields are in need of empirical inputs, which are generally obtained
from experiments or first-principles calculations,
and the prediction power of these force fields requires to
be thoroughly addressed via systematic tests.
To the best of our knowledge, we are not aware of any valid force fields that have been
well tested for studying deuterium diffusion in liquid tin.
Thus, we use first-principles quantum mechanics as a first attempt to tackle this problem.
Density functional theory,\cite{hohenberg64,kohn65} a widely used first-principles method based on quantum mechanics,
has become a powerful tool in predicting properties of various materials.
In particular, density functional theory has been validated to be accurate enough
to study bulk properties of solid tin.\cite{03-Sn-SIESTA}
Furthermore, molecular dynamics simulations based on the density functional theory
have been proven to be suitable for studying liquid tin.\cite{03B-Itami,calderin2008structural}
A previous FPMD study of liquid tin on a 64-atom cell showed
that the computed self-diffusion coefficients of tin atoms were only
about half of the experimental values,\cite{03B-Itami}
implying that the deviation may come from the use of a small cell.\cite{03B-Itami}
The result was later improved in a FPMD simulation of a larger cell consisting of 205 tin atoms,
\cite{calderin2008structural} where the computed diffusivity of liquid tin agrees
well with experiments.

In this work, we perform FPMD simulations
to study deuterium diffusion in liquid tin in a wide range of temperatures (573 to 1673 K).
We begin by validating the ground-state properties solid tin phases.
We then compute the radial distribution functions, static structure factors,
and diffusion coefficients of liquid tin, all of which are in reasonably good agreement with experiments.
Next, we study how the structures and dynamics of liquid tin change upon inserted deuterium atoms.
Our simulations predict that the deuterium atoms in liquid tin diffuse faster than the self-diffusivity of tin.
Additionally, we find that the diffusivity and structures of tin are insensitive to the inserted deuterium at temperatures and concentrations considered.
We also observe that tin and deuterium do not form stable solid compounds.
These predicted results from simulations enable us to have a better understanding of the retention of hydrogen isotopes in liquid tin.
The rest of the paper is organized as follows. We introduce the computational methods in Sec. II.
The results are presented and discussed in Sec. III. We draw conclusions in Sec. IV.

\section{Computational Methods}

The FPMD simulations of tin were performed with the ABACUS
(Atomic-orbital Based Ab-initio Computation at USTC) package.\cite{Li15}
ABACUS was developed for large-scale density functional theory simulations based
on a set of linear combination of atomic orbitals (LCAO).\cite{chen10,chen11}
The plane-wave (PW) basis set is an alternative choice in ABACUS.
By taking the advantage of real-space locality,
the LCAO basis set is more efficient for large systems when compared to the PW basis set.
The recently developed systematically improvable optimized numerical atomic orbitals \cite{chen10,chen11} were
found to be an excellent choice in describing a variety of materials.\cite{Li15}
We adopted the norm-conserving pseudopotentials\cite{giannozzi09} for both tin and deuterium.
The exchange-correlation functionals have different performances for solid tin.
The local density approximation\cite{80L-CA,81B-Perdew} was found to perform better \cite{03-Sn-SIESTA} than the
generalized gradient approximation\cite{96L-PBE} for structural and elastic properties of solid tin,
whereas the latter one provides more accurate binding energies.
For the liquid tin, the structural and dynamical properties of liquid tin
at different temperature were accurately captured by the local density approximation
in previous studies.\cite{03B-Itami,calderin2008structural}
For example, the radial distribution functions, dynamic structure factors, and diffusion coefficients
were found to match reasonably well with experimental data.
Therefore, we decided to use the local density approximation.
The energy cutoff for plane-wave basis set was set to be 16 Ry.
The radius cutoffs of numerical atomic orbitals were chosen to be 8.0 bohr.
The atomic orbitals basis set of tin includes two {\it s}, two {\it p}, and one polarized ({\it d}) orbitals.
We utilized two {\it s} and one polarized ({\it p}) orbitals for deuterium.
A 10$\times$10$\times$10 k-point mesh was adopted to sample the Brillouin zone for solid phases of tin.
The Murnaghan's equation of state\cite{bulk-moduli} was employed to calculate the bulk moduli.
All calculations were performed with periodic boundary conditions.

We performed the Born-Oppenheimer molecular dynamics by utilizing the canonical ensemble NVT
(constant number of particles {\it N}, constant volume {\it V},
and constant temperature {\it T}) with the Nos\'{e}-Hoover thermostat.\cite{nose84,hoover85}
We used the Verlet algorithm and the thermostat mass was chosen to ensure the fluctuations of temperature to be within 0.5\%.
The gamma point was adopted in the {\it k}-point sampling of the Brillouin zone.
The masses of tin and deuterium were chosen to be 118.71 and 2.014 amu, respectively.
In order to ensure the accuracy of integrals performed in Newton's equations of motion for atoms with different masses,
the time step was chosen to be 1.0 fs for pure liquid tin and 0.2 fs when deuterium atoms were inserted.
Two simulation cells were tested in order to validate the size effect\cite{04JPCB-Yeh}: 64- and 216-atom cells.
The liquid densities of tin at different temperatures were chosen based on the experimental data in Ref.~\onlinecite{95PCL-Nasch}.
The liquid tin structure was first prepared by heating the $\alpha$-tin structure at 1073 K for 5.0 ps.
We then ran a 30 ps trajectory at each given temperature between 573 and 1673 K.
Next, we randomly inserted a few deuterium atoms into these 216-atom liquid tin cells.
For each system at a given temperature, the equilibrium process was run for 4 ps followed by a production
trajectory for another 16 ps (80,000 steps).
We ran five concentrations of deuterium at 1073 K in order
to understand the concentration effect on the diffusion coefficients.
Note that we did not change the cell volumes of liquid tin after deuterium atoms were inserted.
The reason is that the averaged pressures on cells are only increased within 5.5 kB (20 deuterium atoms),
which is close to the fluctuation of 5.0 kB observed in our MD trajectories.
Thus, we do not expect the slightly increased pressures to have significant effects on the final results.

We analyze the structural and dynamic properties of liquid tin and liquid tin-deuterium systems through a few tools.
The radial distribution function $g(r)$ is calculated based on
\begin{equation}
g(r) = \frac{1}{\rho N}\langle\sum_{i=1}^{N}\sum_{j=1,j\neq i}^{N}\delta (\mathbf{r} - \mathbf{R}_i + \mathbf{R}_j)\rangle,
\end{equation}
where $\rho$ is the ionic density, $N$ is the total number of atoms,
$\mathbf{R}_i$ and $\mathbf{R}_j$ are atomic coordinates of atoms $i$ and $j$, respectively.
Next, the partial radial distribution function\cite{partial-pdf}
between two species $\alpha$ and $\beta$ can be written as
\begin{equation}
g_{\alpha\beta}(r) = \frac{N}{\rho
N_{\alpha}N_{\beta}}\langle\sum_{i=1}^{N_\alpha}\sum_{j=1}^{N_\beta}\delta(\mathbf{r} - \mathbf{R}_i^{\alpha} + \mathbf{R}_j^{\beta})\rangle.
\end{equation}
The static structure factor $S(q)$ has the form of
\begin{equation}
S(q) = \frac{1}{N}\langle\sum_{i=1}^{N}\sum_{j=1}^{N}e^{i\mathbf{q}\cdot(\mathbf{R}_i-\mathbf{R}_j)}\rangle,
\end{equation}
where $\mathbf{q}$ is the reciprocal space vector and $q=|\mathbf{q}|$.
The formula of diffusion coefficient is
\begin{equation}
D = \frac{1}{6}\frac{d}{dt}\langle\mathrm{\Delta}r(t)^2\rangle,
\end{equation}
where $\mathrm{\Delta}r(t)^2$ is the mean square displacement of atoms at time $t$.
In our study, less accurate statistics are obtained for deuterium than tin atoms
because there are a smaller number
of deuterium atoms than tin atoms in the simulations.
Consequently, the mean square displacements computed for deuterium atoms are
less accurate than those for tin atoms.
In order to obtain trusted diffusion coefficients,
we divided each tin-deuterium trajectory into five segments and calculated the final diffusion coefficient
by averaging the diffusion coefficients computed from all segments.
The diffusion data were shown with the standard deviation.
We further adopt a decay function $H(t)$ to compute the lifetimes of tin-deuterium bonds.\cite{bond-life}
The decay function represents the fraction of unbroken bonds at time $t_m$ and is defined as
\begin{equation}
H(t_m) = \sum_{n=m}^{\infty} N(t_{n+1}) / \sum_{n=1}^{\infty} N(t_n),  H(0) = 1,
\end{equation}
where $N(t_n)$ is the number of bonds that break after $n$ steps.
A bond would be treated as a new one if it breaks and then reforms.
The mean lifetime can be then defined as
\begin{equation}
\tau = \sum_{n=0}^{\infty} \frac{1}{2} \Delta t \Bigl[H(t_n) + H(t_{n+1})\Bigr],
\label{lifetime}
\end{equation}
where $\Delta t$ is the time step.

\section{Results and Discussion}

\subsection{Solid Tin}

\begin{table}[tp]
\caption{Bulk properties of four solid tin structures: $\beta$-tin, $\alpha$-tin, $bct$, and $bcc$.
Lattice constants ($a_0$ and $c_{0}$ in \AA), ratios between two lattice constants ($c_0/a_0$),
volumes ($V_0$ in \AA$^3$/atom), relative energies ($\Delta$$E$ in eV/atom) with the energy of $\beta$-tin set to zero,
and bulk moduli ($B_0$ in GPa) obtained from first-principles (FP) calculations and experiments (EXP).
PW and LCAO refer to FP calculations based on plane-wave basis and linear combination of
atomic orbitals basis sets, respectively.
}
\begin{tabular}{c|cccccc}
\hline
\hline
        & $a_0$ & $c_0$/$a_0$ & $V_0$ & $\Delta$$E$ & $B_0$ & Method \\
\hline

$\beta$-tin  &5.780   & 0.537    & 26.15   &0.000    & 58     & FP (PW) \\
          &5.786     & 0.538    &26.23   &0.000   & 57 & FP (LCAO) \\
          &  5.831    &0.546 & 27.07 & - & - & EXP\cite{66Barrett}   \\
          & 5.8119    &0.543 & 26.65 & - & - & EXP\cite{rayne60}\\
          & - & - & - & - &  57.037  & EXP\cite{Smithells-1983}\\
          & - & - & - & - &  57.9  & EXP\cite{Kennedy}\\
\hline
          $\alpha$-tin   & 6.442  & - & 33.41 & -0.019   & 43 & FP (PW) \\
              & 6.445 & - & 33.47 & -0.055   & 43 & FP (LCAO) \\
              & 6.483 & - & 34.05 & - & - & EXP\cite{66Barrett}  \\
              & - & - & - & - & 42.617 & EXP\cite{Smithells-1983}  \\
              & - & - & - & - & 54 & EXP\cite{71B-Pollack} \\
\hline
bct       & 3.933     & 0.846   &25.73   &0.045 & 53  & FP (PW) \\
          &3.920      & 0.844   &25.42   &0.047 & 54  & FP (LCAO)  \\
          \hline
bcc       &3.664      &     &24.60    &0.095  &70   & FP (PW) \\
          &3.658      &     &24.47    &0.130  &71   & FP (LCAO)  \\
\hline
\hline

\end{tabular}
\label{tab:solid}
\end{table}

We benchmark four phases of solid tin
including $\alpha$-tin, $\beta$-tin, body-centered tetragonal ($bct$), and body-centered cubic ($bcc$).
Experimentally, the $\alpha$-tin phase has the lowest energy, and transforms to
$\beta$-tin with a tetragonal crystal structure at 286.3 K and atmospheric pressure.\cite{35Cohen}
At room temperature, the $\beta$-tin structure transforms to the $bct$ structure
at 9.5 GPa\cite{84-Olijnyk} and $bcc$ at approximately 45 GPa,\cite{89B-Desgreniers}
and the $bcc$ phase is stable up to 120 GPa.\cite{89B-Desgreniers}
Table~\ref{tab:solid} lists the first-principles results and available experimental data.
We calculated several bulk properties including lattice constants, equilibrium volumes,
energy orderings, and bulk moduli.
Note that PW and LCAO basis sets were used in calculations
and results from both basis sets agree well with each other.
For instance, the $c_{0}/a_{0}$ ratio of $bct$ (0.844) from the LCAO basis only differs
within 0.3\% when compared to the one (0.846) from the PW basis.
Although the computed lattice constants of $\alpha$-tin and $\beta$-tin are slightly smaller (within 0.9$\%$)
than the experimental values, this level of discrepancies is expected due to
the use of the local density approximation\cite{80L-CA,81B-Perdew} which tends to yield stronger binding between atoms.
We also find that the bulk moduli calculated from the two basis sets are all close within 1 GPa
and agree well with experiments.
Notably, both basis sets yield the same energy orderings among
the selected four phases of solid tin: $\alpha$-tin is the most stable phase, followed by
$\beta$-tin, $bct$, and $bcc$ phases.
However, the LCAO basis predicts a 36 meV/atom of larger energy difference between $\alpha$-tin and $\beta$-tin than the PW basis does.
The error probably comes from the incompleteness of the LCAO basis set.
Although this energy difference may be important in calculating the phase transition between $\alpha$-tin and $\beta$-tin,
we demonstrate that it does not affect our simulations of liquid tin, as evidenced in the following section.
Overall, we find that the first-principles calculations with the LCAO basis set
well capture the bulk properties of solid tin crystals.

\subsection{Liquid Tin}


\begin{figure}[htbp]
\centering
\includegraphics[width=0.42\textwidth, clip]{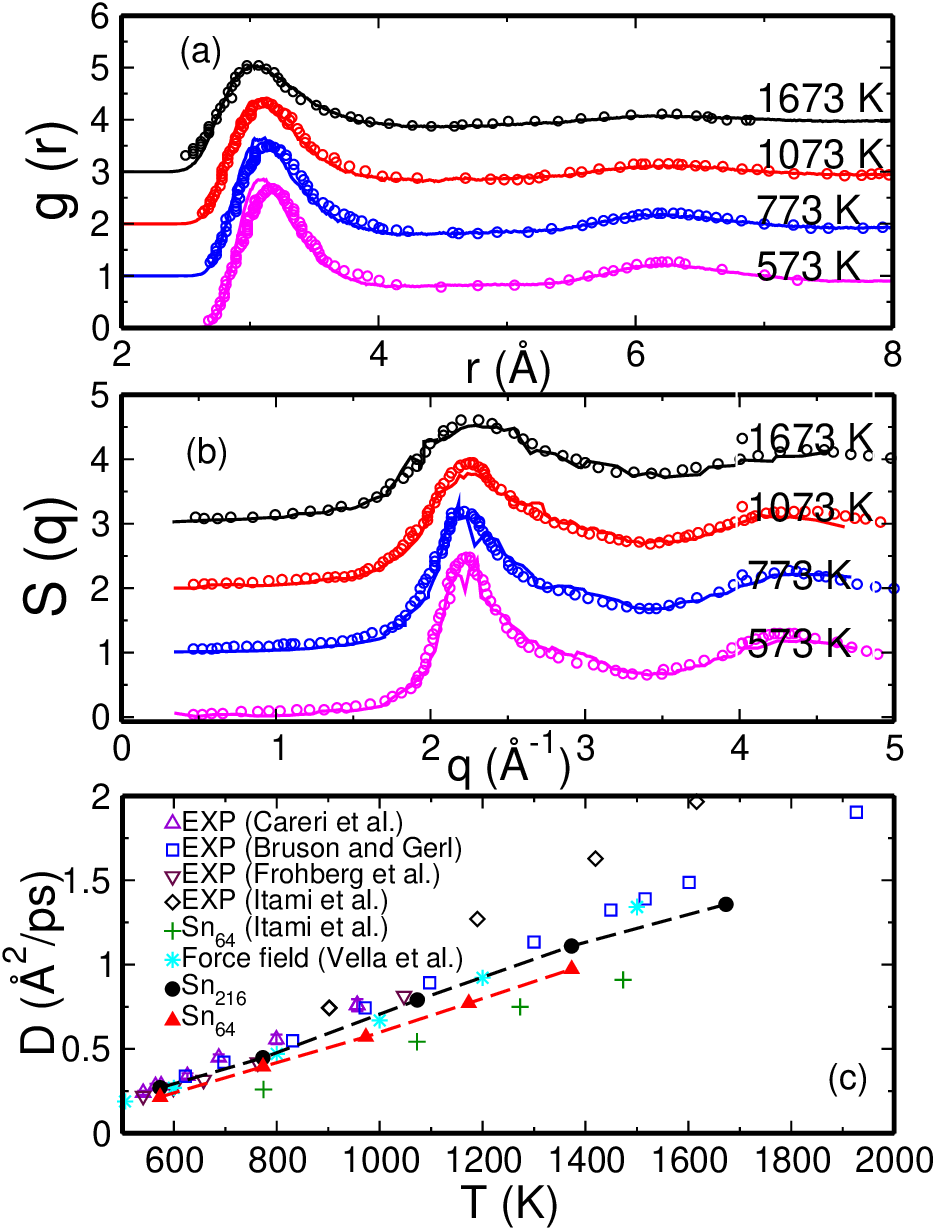}
\caption{(a) Radial distribution functions $g(r)$ and (b) static structure factors $S(q)$
obtained from first-principles molecular dynamics simulations of liquid Sn (216 atoms) at temperatures 573, 773, 1073, and 1673 K.
First-principles and experimental results (EXP)\cite{03B-Itami} are depicted in solid lines and circles, respectively.
The $g(r)$ and $S(q)$ data at temperatures other than 573 K are shifted upwards for ease of viewing.
(c) Diffusion coefficients of Sn in the 216-atom (black) and 64-atom cells (red) are compared to
the experimental results from Careri et al.,\cite{Careri} Bruson and Gerl,\cite{Bruson}
Frohberg  et al.,\cite{Frohberg} and Itami et al.\cite{03B-Itami}
The first-principles molecular dynamics results from Itami et al.\cite{03B-Itami} using a 64-atom cell are also included.
We also include the data (cyan) computed by classical force field from Ref.~\onlinecite{17B-Vella}.
}
\label{fig:sq}
\end{figure}

\begin{table}[t]
\caption{Peak positions ($r_1$ and $r_2$) of the radial distribution function $g(r)$ of liquid Sn at different temperatures.
Experimental data (EXP) are taken from Itami, et al.\cite{03B-Itami}
First-principles molecular dynamics (FPMD) data are obtained from the LCAO basis set in a Sn cell consisting of 216 atoms.
$g(r_1)$ and $r_1$ depict the height and position of the first peak, while
$g(r_2)$ and $r_2$ depict the height and position of the second peak.
Two ratios $R_r=r_2/r_1$ and $R_g=g(r_2)/g(r_1)$ are listed. $N_c$ is the
first coordination number of liquid Sn.}
\begin{tabular}{c|cccccccc}
\hline
\hline
  $T$ (K) & Method  & $r_1$ & $g(r_1)$ & $r_2$ & $g(r_2)$ & $R_r$ & $R_g$  & $N_{c}$ \\
\hline
573 & EXP &3.169      &2.697     &6.241     &1.274    &1.97   &0.472  &10.7\\
    & FPMD &3.118      &2.783     &6.178     &1.243    &1.98   &0.447  &10.5\\
773 & EXP &3.149      &2.527     &6.253     &1.217    &1.99   &0.482  &10.2\\
    & FPMD &3.102      &2.525     &6.162     &1.142    &1.99   &0.452  &10.1\\
1073& EXP &3.121      &2.347     &6.188     &1.162    &1.98   &0.495  &9.6\\
    & FPMD &3.071      &2.320     &6.162     &1.091    &2.01   &0.470  &9.7\\
1673 & EXP &3.099 &2.086 &6.354 &1.104 &2.05 &0.529   & 8.6    \\
     & FPMD &3.085 &2.022 &6.095 &1.076 &1.98 &0.532   & 8.8    \\
\hline
\hline
\end{tabular}
\label{tab2}
\end{table}

We next compare several properties of liquid tin obtained from simulations to available experiments.
Figures 1(a) and (b) illustrate the radial distribution functions $g(r)$ and
static structure factors $S(q)$ of liquid tin, respectively.
The experimental data of $S(q)$ were obtained from neutron scattering experiments.\cite{03B-Itami}
We show $g(r)$ and $S(q)$ at temperatures ranging from 573 to 1673 K
except the ones at 1373 K because the experimental data were suggested to be inaccurate.\cite{03B-Itami}
In general, we find that the predicted shapes and positions of peaks in both $g(r)$ and $S(q)$ at these four temperatures
match very well with the experiment data.
For instance, the computed first and second peaks of $g(r)$ and $S(q)$ decrease with elevated temperatures,
which agree with the experiment and suggest more liquid-like structures of liquid tin at higher temperatures.
In order to quantitatively compare the simulation data to experiments,
we list more characteristic features of radial distribution functions in Table II, which contains
the positions ($r_1$, $r_2$) and heights ($g(r_1)$, $g(r_2)$) of the first and second peaks.
Table II also includes the ratio ($R_r$) between $r_1$ and $r_2$,
the ratio ($R_g$) between $g(r_1)$ and $g(r_2)$, and the first coordination number $N_{C}$.
The data shown in Table II agree quite well with the experiment.\cite{03B-Itami}
For example, the computed $N_{C}$ decreases from 10.5 (573 K) to 8.8 (1673 K),
which is consistent with the experimental value that decreases from 10.7 (573 K) to 8.6 (1673 K).
In addition, both $R_r$ and $R_g$ from simulations are close to the experimental data except at 1673 K.
We point out that some discrepancies exist in $r_2$ and $R_r$ at 1673 K because
the smooth second peak of $g(r)$ at the temperature was difficult to be located.
We also notice that both $r_1$ and $r_2$ are reasonably accurate but slightly smaller than the experimental data.
These discrepancies can be attributed to the utilization of the local density approximation that overbinds atoms.
Furthermore, the shoulder feature of $S(q)$ on the high-$q$ side of
the first peak (around 2.8 \AA$^{-1}$) is captured by our simulations,
in consistent with a previous first-principles study by Itami et al.\cite{03B-Itami}
More detailed discussion on the shoulder feature of $S(q)$ can be found in Ref.~\onlinecite{03B-Itami}.
Note that there are some noisy features appear on the computed $S(q)$ in Figure 1(b),
which are due to the fact that static structure factors are more difficult to converge
than the radial distribution functions.\cite{03B-Itami}

Figure 1(c) illustrates the diffusion coefficients of tin in both 64- and 216-atom cells,
as well as four sets of available experimental data.\cite{Careri,Bruson,Frohberg,03B-Itami}
The experimental data are in excellent agreement at temperatures lower than 1200 K,
so our comparison is mainly discussed in this temperature range.
We observe that our diffusion coefficients of tin (red) from the 64-atom cell
are close to those first-principles data (green) of Itami et al.,\cite{03B-Itami}
albeit both are significantly smaller than the experimental values.
The underestimation of the diffusion coefficients is probably due to the size effect,
because the imposed periodic boundary conditions create artificial
interactions between images of liquid atoms, causing large errors in small cells.\cite{03B-Itami,Chen15}
As expected, the diffusion coefficients calculated from the 216-atom cell (shown in black dots)
are larger than those obtained from the 64-atom cell and in better agreement with experimental data.
Although the size effect is still expected in the 216-atom cell and should
be clarified in future studies, we consider the cell size is a good balance
between accuracy and efficiency.
In this regard, we study the deuterium diffusion in liquid tin by adopting the 216-atom cell.

\subsection{Deuterium in Liquid Tin}
To the best of our knowledge, we cannot find any FPMD simulations of deuterium diffusion in liquid tin reported.
One of the reasons is that performing first-principles
simulations on cells containing hundreds of atoms at different temperatures
is computationally demanding.
With the aid of numerical atomic orbitals as accurate basis sets,
we are able to reduce the computational cost.
We first validated the basis set for hydrogen (deuterium) atoms, where
calculations within the framework of density functional theory gave rise to the same structural properties
including the bond length and formation energy because the mass of the nuclei did not play a role.
The computed bond lengths of hydrogen dimer are 0.766 \AA~for PW basis set
compared with 0.761 \AA~for LCAO basis set, which are close to the experimental
length of 0.74 \AA.\cite{dekock1989chemical}
The binding energies of hydrogen dimer are 4.85 and 4.95 eV from PW and LCAO
basis sets, respectively;
the computed binding energies are somehow larger than the experimental value of 4.467 eV~\cite{dekock1989chemical}
which are consistent with the overbinding feature of local density approximation.
Next, we obtained diffusivity data
and structural properties of liquid Sn$_{1-x}$D$_{x}$ by using molecular dynamics.
These results include the temperature- and concentration-dependent diffusion coefficients of deuterium and tin,
and the bonding status between deuterium and tin.

\begin{figure}[tbp]
\centering
\includegraphics[width=0.4\textwidth, clip]{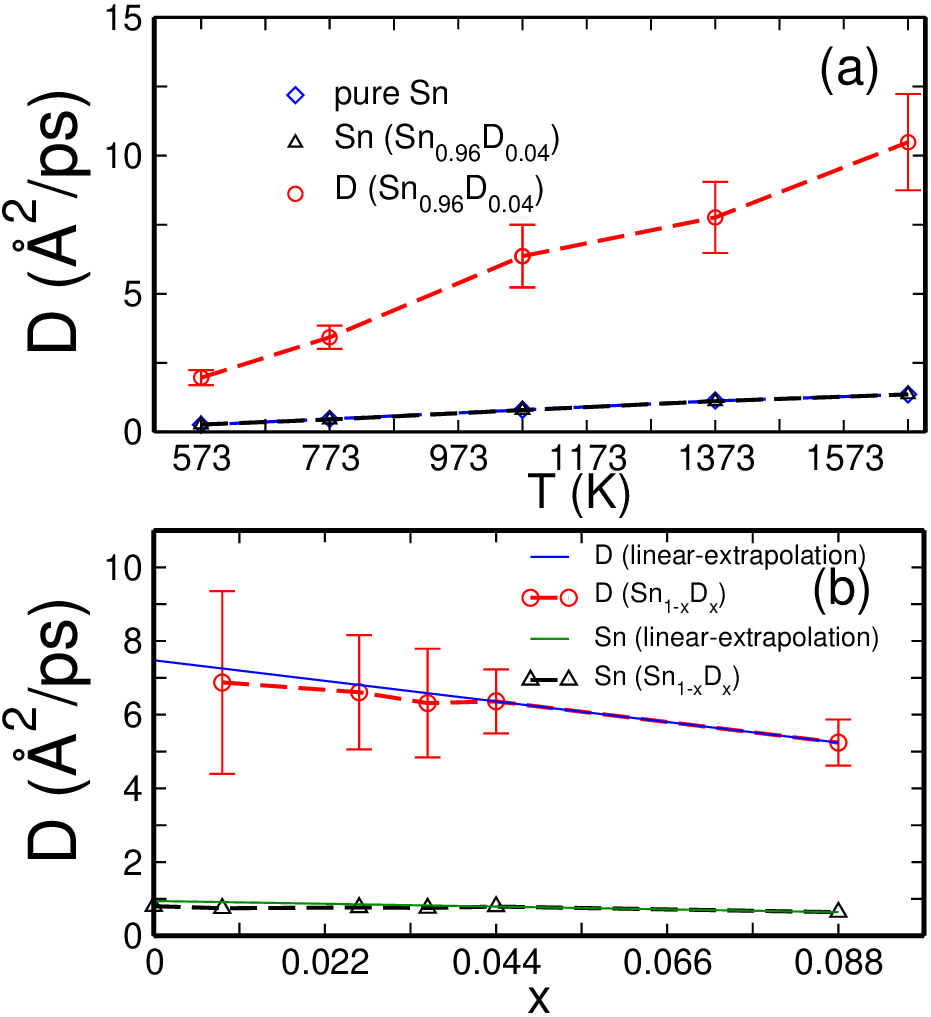}
\caption{(a) Diffusion coefficients of D (red circles) in liquid Sn$_{0.96}$D$_{0.04}$, Sn (black triangles)
in Sn$_{0.96}$D$_{0.04}$, and Sn (blue diamonds) in pure liquid Sn at 573, 773, 1073, 1373, and 1673 K.
(b) Diffusion coefficients of D (red circles) and Sn (black triangles) in liquid Sn$_{1-x}$D$_{x}$ at 1073 K with
$x$ being 0.009, 0.027, 0.036, 0.044, and 0.085. The diffusion coefficient of pure liquid Sn ($x$=0.0) at 1073 K is also shown.
We also include linear extrapolation data for deuterium (solid blue line) and tin (solid green line) atoms based on
the diffusion data obtained from systems with $x=$0.044 and 0.085.}
\label{fig:diffusion}
\end{figure}

Figure 2(a) shows the diffusion coefficients of deuterium (10 atoms) in a 216-atom liquid tin cell
at five temperatures ranging from 573 to 1673 K.
We find that the diffusion coefficient of deuterium increases faster than that of tin as temperature increases.
For instance, the diffusion coefficients of deuterium are 7.3 and 7.7 times larger than those of tin at 573 and 1673 K, respectively.
In addition, we observe that the diffusion coefficients of pure liquid tin almost remain the same values even
with the presence of inserted deuterium atoms
at all temperatures considered, in stark contrast to the diffusion coefficients of lithium that can be significantly
affected with the presence of deuterium.\cite{16NF-Chen}

Figure 2(b) illustrates the impact of the concentration of deuterium on the diffusivities of both deuterium and tin atoms at 1073 K.
We find that the concentrations of deuterium, varying from 8.47\% (20 deuterium atoms in a 216-atom liquid tin cell)
to 0.92\% (2 deuterium atoms in a 216-atom liquid tin cell),
have relatively smaller impacts on the diffusion coefficients of deuterium as compared to the temperature effect in our simulations.
Specifically, we observe that the diffusion coefficient of deuterium increases from 5.2 to 6.9 ${\rm \AA^2/ps}$
when its concentration decreases from 8.47\% to 0.92\%.
However, the diffusion coefficient of deuterium (concentration is 4.42\%)
largely increases from 2.0 to 10.5 ${\rm \AA^2/ps}$ when temperature is elevated from 573 to 1673 K as shown in Figure 2(a).
Based on the temperatures (573 to 1673 K) and concentrations of deuterium (0.92\% to 8.47\%) considered in our simulations,
we suggest that temperatures impact the diffusion coefficients of
deuterium more significantly than concentrations of deuterium.
We notice that the error bars of the diffusion coefficients
also increase as the concentration of deuterium decreases, which can be
attributed to the small samplings of deuterium atoms.
Unfortunately, the high computational cost limits our further study of diffusion coefficients of deuterium
at concentrations smaller than 0.92\%, in which cases larger cells are needed.
We also tried linear extrapolations based on the diffusion data obtained
when $x=$0.044 and 0.085, as shown in Figure 2(b). The extrapolated data for both deuterium and tin atoms are
close to the values we obtained by molecular dynamics simulations, supporting the above conclusions.
We further compare the diffusion coefficients of tin at different concentrations of deuterium
to those of pure liquid tin, and find that different concentrations of deuterium
do not affect the diffusion coefficients of tin at the same temperature.
This observation further supports our previous finding
that the diffusion coefficients of tin are not affected by the presence of deuterium at different temperatures considered (573 to 1673 K).

\begin{figure}[htbp]
\centering
\includegraphics[width=0.38\textwidth, clip]{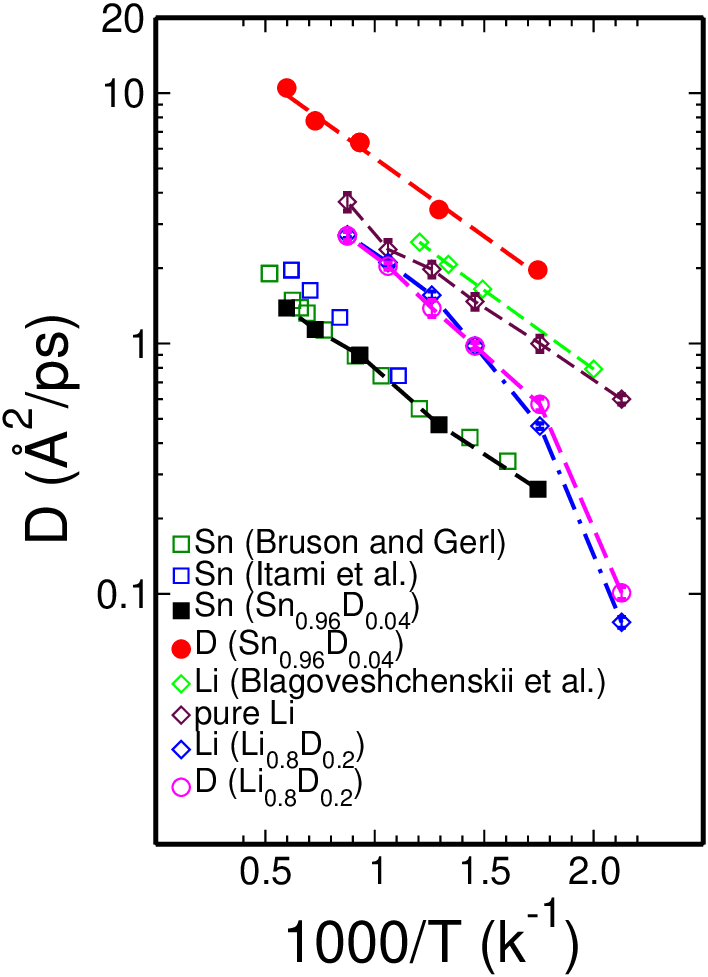}
\caption{Diffusion coefficients of D (red circles) and Sn (black squares) in liquid Sn$_{0.96}$D$_{0.04}$
from first-principles molecular dynamics (FPMD) simulations.
The red dash line is fitted from the diffusion coefficients of D in liquid Sn$_{0.96}$D$_{0.04}$
using Eq.~\ref{fitting}.
Diffusion coefficients of Li (purple diamonds) in pure liquid Li,
D (magenta circles) and Li (blue diamonds) in Li$_{0.8}$D$_{0.2}$
are from FPMD simulations as done in Ref.~\onlinecite{16NF-Chen}.
Experimental diffusion coefficients include pure Li (green diamonds) from Blagoveshchenskii et al.,\cite{blagoveshchenskii2012self}
pure Sn (blue squares) from Itami et al.,\cite{03B-Itami} and pure Sn (dark green squares) from Bruson and Gerl.\cite{Bruson}
}
\label{fig:fit}
\end{figure}

Although only limited data are available,
it is informative to compare the behaviors of deuterium in liquid tin and liquid lithium
at different temperatures.
We summarize the diffusivity data in Figure 3 including both experimental and available first-principles simulation data.
Specifically, the first-principles results of lithium and deuterium in the Li$_{0.8}$D$_{0.2}$ system are taken from Ref.~\onlinecite{16NF-Chen},
where simulations were performed in a cell containing 128 lithium and 32 deuterium atoms;
data in terms of higher concentrations of deuterium in lithium are not included here,
but the trend is that the diffusivities of both deuterium and lithium at a given temperature
decrease as the concentration of deuterium increases.
The decrease of diffusion coefficients is due to the formation of solid lithium deuteride,
which has a high melting point of 965$\pm$2 K.\cite{messer1965systems}
However, we did not observe formation of any solid compounds in our simulated tin-deuterium systems.
As shown in Figure 3 with a dashed red line, we fit the computed temperature-dependent diffusion coefficients of deuterium
in Sn$_{0.96}$D$_{0.04}$ into an equation as\cite{moriyama1992transport,fukada2005hydrogen}
\begin{equation}
D(T) = 23.30 \times \mathrm{exp} (- 11980~[\mathrm{J}\cdot \mathrm{mol}^{-1}]/RT)~[{\rm \AA^2/ps}],
\label{fitting}
\end{equation}
where $R$ is the gas constant (in J$\cdot$mol$^{-1}$$\cdot$K$^{-1}$) and $T$ (in K) is temperature.
Besides the simulation data, we also show experimental data including lithium
diffusivities from Blagoveshchenskii et al.,\cite{blagoveshchenskii2012self}
and tin diffusivities from Itami et al.\cite{03B-Itami} (blue squares) and Bruson and Gerl (dark green squares).\cite{Bruson}
There are a few interesting observations in Figure 3.
First, both diffusivities of pure lithium and tin are accurately captured by
the FPMD simulations when compared to available experiments,
suggesting that simulations are able to yield trustable diffusion coefficients.
Second, the diffusivity of lithium in Li$_{0.8}$D$_{0.2}$ is smaller than that of pure lithium,
and the difference becomes more significant at relatively lower temperatures.
For example, the diffusion coefficients of lithium and deuterium
significantly drop by about one order of magnitude from 570 to 470 K,
which can be explained by the formation of strong chemical bonds between lithium and deuterium.\cite{16NF-Chen}
Third,
we find that the diffusivity of tin is insensitive to the presence of inserted deuterium (concentration from 0.92\% to 8.47\%)
in a wide range of temperatures (573 to 1673 K) considered here.
Finally, the deuterium in liquid tin diffuses faster than both pure tin and pure lithium.

\begin{figure}[tbp]
\centering
\includegraphics[width=0.4\textwidth, clip]{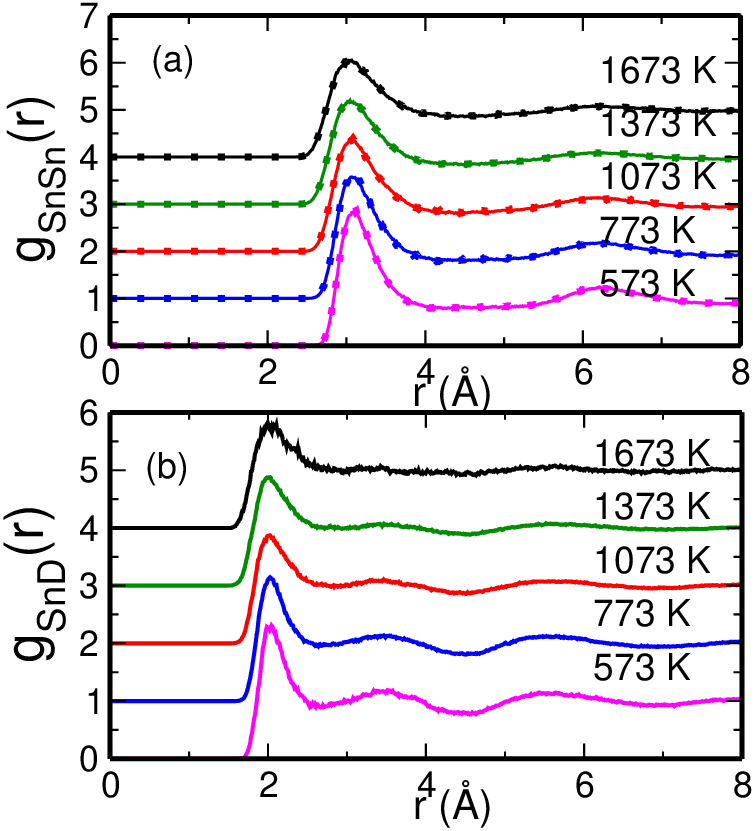}
\caption{Partial radial distribution functions of (a) $g_{\mathrm{SnSn}}(r)$ (solid lines) and (b) $g_{\mathrm{SnD}}(r)$ (solid lines)
 of liquid Sn$_{0.96}$D$_{0.04}$ at 573, 773, 1073, 1373, and 1673 K.
 The dotted lines in (a) represent $g(r)$ of pure liquid Sn from first-principles molecular dynamics simulations.
 $g_{\mathrm{SnSn}}(r)$ and $g_{\mathrm{SnD}}(r)$ at temperatures other than 573 K are shifted upwards for ease of viewing.
}
\label{fig:pdf-SnD-tem}
\end{figure}
\begin{figure}[t]
\centering
\includegraphics[width=0.42\textwidth, clip]{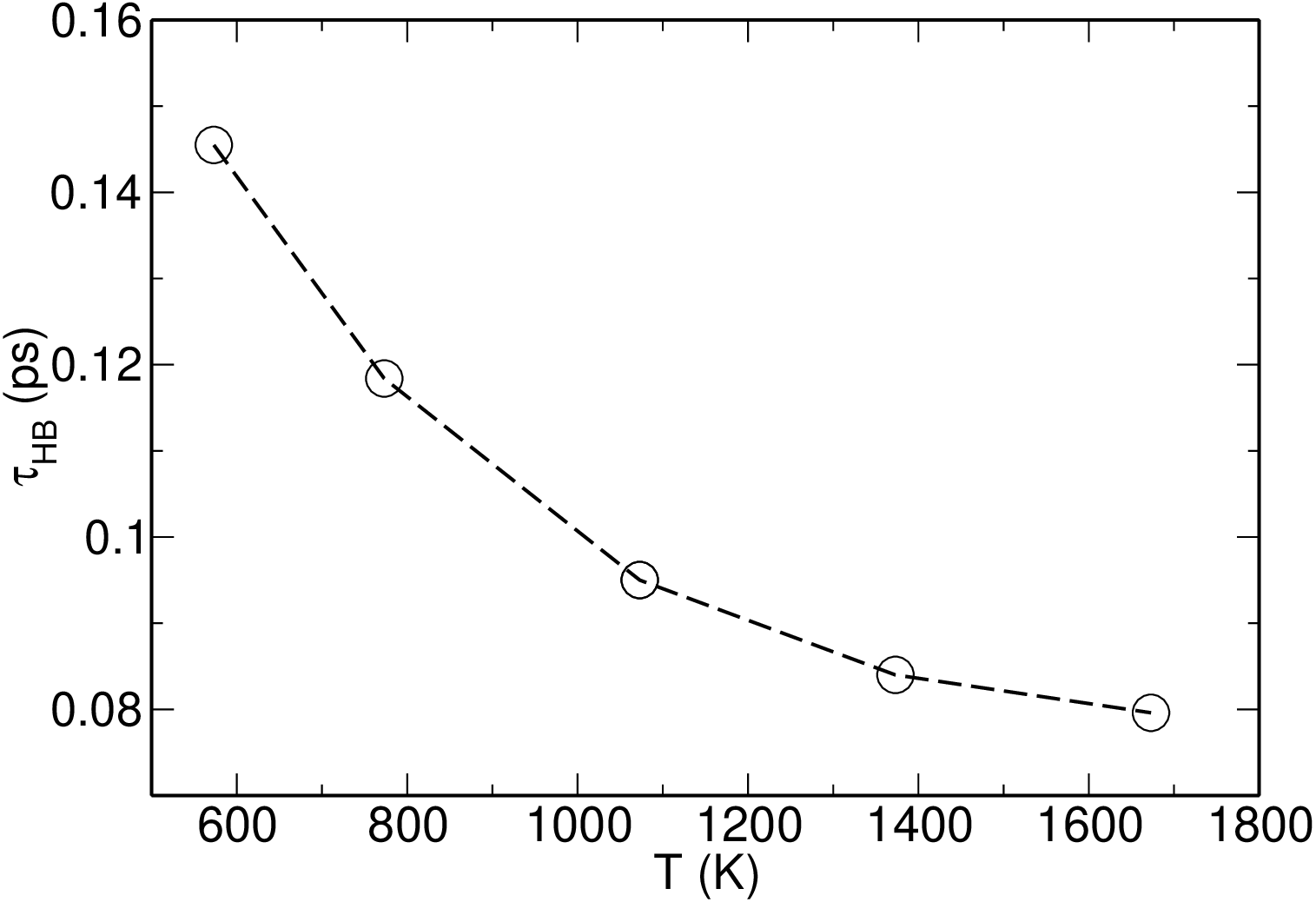}
\caption{Lifetimes of the Sn-D bond in liquid Sn$_{0.96}$D$_{0.04}$  at 573, 773, 1073, 1373, and 1673 K.
}
\label{fig:SnD}
\end{figure}

The different behaviors of deuterium predicted in liquid tin and lithium
can be partially attributed to the different chemical bonding statuses between deuterium and liquid metals.
In this regard, we calculated the partial radial distribution functions
and performed lifetime analysis to obtain bonding information between deuterium and tin in liquid Sn$_{0.96}$D$_{0.04}$.
Figures 4(a) and (b) show the partial radial distribution functions
of tin-tin ($g_{\mathrm{SnSn}}(r)$) and tin-deuterium ($g_{\mathrm{SnD}}(r)$)
at five temperatures ranging from 573 to 1673 K.
Two interesting features are discussed here.
First, the positions and heights of the first and second peaks of $g_{\mathrm{SnSn}}(r)$ (solid lines)
in Figure 4(a) are almost indistinguishable to those $g(r)$ (dotted lines) in
pure liquid tin (Figure 1(a)) at each given temperature, indicating that deuterium atoms
have only little impact on the liquid structure of tin atoms.
Second, the position of the first peak of $g_{\mathrm{SnD}}(r)$ is predicted to be around 2.1 \AA, which is
shorter than that of $g_{\mathrm{SnSn}}(r)$ (about 3.1 \AA).
This implies that deuterium atoms prefer to stay between tin and its first shell of tin neighbors.
In addition, the first minimum of $g_{\mathrm{SnD}}(r)$ is at 2.6 \AA~and we chose it
as a cutoff for lifetime analysis.
Figure 5 illustrates the calculated average lifetimes (Eq.~\ref{lifetime}) of the tin-deuterium bonds at different temperatures.
We can see that the lifetime of the tin-deuterium bonds monotonically decreases as temperature increases.
More importantly, the largest lifetime is 0.145 ps for the tin-deuterium bond at 573 K,
implying that tin and its adjacent deuterium atoms dissociate quickly.
We also observe formation of a few
deuterium pairs that last for 0.1 ps on average,
but the pairs dissociate quickly.
We do not observe formation of deuterium clusters in liquid tin with inserted deuterium.

\section{Conclusions}
Liquid tin is a promising candidate of liquid metal plasma-facing materials that have
substantial advantages over solid materials.
Therefore, the interactions between liquid tin and the fuels of a fusion reactor, i.e., hydrogen isotopes
should be thoroughly understood.
For example, the knowledge of retention and recycling of hydrogen isotopes in liquid metal at different temperatures
should be well documented.
However, the diffusion coefficients of hydrogen isotopes
in liquid metals such as lithium and tin are still controversial or even lacked from experiments.
Classical force fields could be an ideal tool to study deuterium diffusion in liquid tin, but
such models have not yet been established.

In this regard,
we have performed first-principles molecular dynamics simulations
to study deuterium diffusion in liquid tin.
We first tested pure liquid tin at temperatures ranging from
573 to 1673 K. The computed radial distribution functions and static structure factors
match reasonably well with the experimental data. Importantly, the diffusion coefficients
calculated from a 216-atom cell agree better with experiments than those from a 64-atom cell,
suggesting the 216-atom cell is more suitable for studying diffusivities of liquid tin.
Next, we studied liquid tin with inserted deuterium and predicted
the diffusion coefficients of deuterium in liquid tin at different temperatures.
We predicted several interesting results. First, we found faster
diffusion of deuterium in liquid tin than the self-diffusion of tin.
Second, the tin diffusivity and structures are insensitive to the inserted deuterium at temperatures and concentrations considered\MC{.}
Finally, tin and deuterium do not form stable tin-deuterium solid compounds in liquid Sn$_{1-x}$D$_{x}$.
These predicted results from first-principles molecular dynamics not only gave us
a better understanding of the retention of hydrogen isotopes in liquid tin,
but can also be used to develop classical tin-deuterium force fields for large scale simulations.

In future, quantum-mechanics-based first-principles simulations and
classical force fields can be utilized as important computational tools in understanding more
fundamental properties of plasma-facing materials and their interactions with hydrogen isotopes and impurities
at different environments.
For example, tritium atoms can be introduced into liquid metals and their isotope effects can be studied.
Moreover, it is worth examining the diffusion of hydrogen isotopes in liquid lithium-tin mixtures, which
are also candidates of liquid metal plasma-facing materials.

\section*{Acknowledgment}
The authors are grateful to Tyler Abrams and G. G. van Eden for providing valuable input to the manuscript.
This work was funded by the Chinese National Science Foundation (Grant number 11374275),
the National Key Research and Development Program of China (Grants No. 2016YFB0201202).
The numerical calculations have been done on the USTC HPC facilities.

\bibliography{references_Sn}

\end{document}